\documentclass[a4paper,11pt]{article}
\usepackage{pos}
\usepackage{amsmath,graphicx,multirow,tabularx,physics}
\usepackage{davitex,slashed}

\newcommand{\muCEP}{\mu_B^{\text{CEP}}}
\newcommand{\TCEP}{T^{\text{CEP}}}

\title{Searching for the QCD critical point using Lee-Yang edge singularities}
\ShortTitle{QCD critical point from LYE}
\author*[a]{D. A. Clarke}
\author[b]{P. Dimopoulos}
\author[b]{F. Di Renzo}
\author[c]{J. Goswami}
\author[d]{C. Schmidt}
\author[d]{S. Singh}
\author[e]{K. Zambello}

\affiliation[a]{Department of Physics and Astronomy, University of Utah,\\
Salt Lake City, Utah  84112, United States}

\affiliation[b]{Dipartimento di Scienze Matematiche, Fisiche e Informatiche, \\
Università di Parma and INFN, Gruppo Collegato di Parma 
I-43100 Parma, Italy}

\affiliation[c]{RIKEN Center for Computational Science,\\
Kobe 650-0047, Japan}

\affiliation[d]{Universit\"at Bielefeld, Fakult\"at f\"ur Physik,\\
D-33615 Bielefeld, Germany}

\affiliation[e]{Dipartimento di Fisica dell’Università di Pisa and INFN--Sezione di Pisa,\\ 
Largo Pontecorvo 3, I-56127 Pisa, Italy.}

\abstract{
Using $N_f=2+1$ QCD calculations at physical quark mass and purely imaginary 
baryon chemical potential, we locate Lee-Yang edge singularities in the complex 
chemical potential plane. These singularities have been obtained by the 
multi-point Padé approach applied to the net baryon number density. 
We recently showed that singularities extracted with this approach are consistent with universal scaling near the Roberge-Weiss transition.
Here we study the universal scaling of these 
singularities in the vicinity of the QCD critical endpoint. Making use of an 
appropriate scaling ansatz, we extrapolate these singularities on $N_\tau=6$ and $N_\tau=8$ 
lattices towards the real axis to estimate the position of a possible QCD critical point. 
We find an approach toward the real axis with decreasing temperature. 
We compare this estimate with a HotQCD estimate obtained from poles of a 
[4,4]-Padé resummation of the eighth-order Taylor expansion of the QCD pressure.
}

\FullConference{%
 The 40th International Symposium on Lattice Field Theory, LATTICE2023
  31st July - 4th Aug, 2023
  Batavia, Illinois, USA
}

\begin{document}
\maketitle

\section{Introduction}\label{sec:intro}

At sufficiently high $T$ and/or $\mu_B$, nucleons dissociate into their constituents, 
and nuclear matter changes phase from a gas of hadrons to quark-gluon plasma.
At low enough temperature, one eventually encounters with increasing $\mu_B$ a
first-order line, which is expected to terminate at a second-order critical 
endpoint (CEP) belonging to the 3-$d$, $\Z_2$ universality class.
Unfortunately at $\mu_B>0$, the Boltzmann factor becomes complex, 
and a direct estimate of the path 
integral through MCMC is no longer possible--this is the infamous sign problem.
This limitation is a special hindrance when looking out for the possible CEP
at nonzero $\mu_B$ mentioned above. 

Nevertheless many approaches have been developed to at least partially 
circumvent this limitation, 
each with its own merits, drawbacks, and regions of applicability.
One possibility is to carry out simulations at pure imaginary
$\mu_B$, where there is no sign problem. In this approach, observables calculated
on the imaginary axis are then analytically continued to
$\mu_B\in\R$~\cite{deForcrand:2002hgr,DElia:2002tig}.
Another common tactic is to expand the pressure $P$ in
$\mu_B/T$ about $\mu_B/T=0$~\cite{Allton:2005gk,Gavai:2003mf}. Both of these methods have been
successfully employed to, for example, extract higher-order Taylor coefficients
of $P$; unfortunately these quantities become prohibitively computationally
expensive as the order increases, and state-of-the-art calculations
only make it to order 8~\cite{Borsanyi:2018grb,Bollweg:2022fqq}.

These difficulties motivate approaches that allow one to glean more information
about the QCD phase diagram with fewer Taylor coefficients.
The approach we study in these proceedings is the multi-point 
Padé~\cite{Dimopoulos:2021vrk},
which allows one to construct high-order rational approximations
using few Taylor coefficients. This approach has opened up a vibrant
research campaign, facilitating our exploration of the
QCD phase diagram at nonzero chemical 
potential~\cite{Schmidt:2022ogw,Schmidt:2023jcv,Clarke:2023noy}.

\section{Using rational approximations to locate the CEP} 

The Lee-Yang theorem~\cite{Yang:1952be} tells us that the
zeroes of the partition function $\ZQCD$
that approach the real axis as the physical volume $V\to\infty$ correspond to phase
transitions; hence the pressure $P=(T/V)\log\ZQCD$ is expected to be singular
at a phase transition in the thermodynamic limit, and correspondingly, the
closest singularity to the origin in the complex $\mu_B$ plane limits
the radius of convergence of the pressure series.

The CEP is thought to belong to the 3-$d$ Ising universality class.
In that context there are two relevant couplings $t\sim T-\Tc$ and $h$,
the reduced temperature and symmetry-breaking coupling respectively,
that parameterize the distance from the critical point.
Away from $\Tc$, the Lee-Yang singularity gets demoted to
Lee-Yang edges (LYE), two branch cuts in the complex $h$-plane that
pinch the real axis as $T\to\Tc$~\cite{Fisher:1978pf}. According to the extended
analyticity conjecture~\cite{fonseca_ising_2003}, 
the LYE are the closest singularities to the origin.
So the thinking goes that the convergence radius of the pressure
is limited by the LYE corresponding to the CEP\footnote{This statement of course depends on
the nearest phase transition. At high temperatures one needs to consider
the Roberge-Weiss transition, and for small quark masses, the chiral
phase transition has increasing importance.}.
Switching from $h$ to $z\equiv th^{-1/\beta\delta}$, with $\beta$
and $\delta$ being universal critical exponents,
the LYE is known to be located at
\begin{equation}\label{eq:zscaling}
    z_c=|z_c|e^{\pm i\pi/2\beta\delta}.
\end{equation}

Due to the sign problem, lattice calculations cannot be performed
at real $\mu_B$ directly. But \equatref{eq:zscaling} opens up the possibility of
carrying out simulations at pure imaginary $\mu_B$, following the
scaling to learn something\footnote{Assuming we know the mapping
from $T$ and $\mu_B$ to $t$ and $h$.} about real $\mu_B$. Since a phase transition
must occur at pure real $\mu_B$, one can follow the imaginary part
until it hits zero, signalling the transition, and see what the real
part is at that point. This requires us to find complex singularities.

Rational approximations are good candidates for approximating functions with
singularities, as zeroes in their denominator mimic singular behavior.
Given a generic function $f$ of some variable $x$, 
we construct a rational approximation
\begin{equation}
  R_n^m(x)\equiv\frac{\sum_{i=0}^m a_ix^i}{1+\sum_{j=1}^nb_jx^j}.
\end{equation}
To connect the approximation closely to $f$, let $f$ have a formal Taylor series
\begin{equation}
  f(x)=\sum_{k=0}^\infty c_kx^k.
\end{equation}
The Padé approximant of order $[m,n]$ is the rational approximation
$R_n^m$ with coefficients chosen such that $R_n^m$ equals the Taylor series 
up to order $m+n$. This constraint yields a set of equations relating 
coefficients $a_i$, $b_j$, $c_k$.
Some properties of Padé approximants are rigorously known;
for instance a Padé approximant is unique when it exists, and the
$[m,n]$ approximant converges to $f$ exactly as $m\to\infty$ when $f$ has pole
of order $n$. Other properties must be deduced from numerical experiments;
for instance if $f$ has a branch cut, zeroes of the denominator tend
to accumulate along the cut.
Turning to $P$, we are faced with the difficulty mentioned in
\secref{sec:intro} that expansion coefficients are known only up to eighth order,
which therefore limits the order of the corresponding Padé.

To mitigate this difficulty, we use the multi-point Padé. The problem to solve is
that we only have information about the Taylor expansion to low order,
say order $s-1$. The multi-point Padé is a rational 
function $R_n^m$ satisfying for $0\leq l<s-1$
\begin{equation}\label{eq:mpp}
\dv[l]{R^m_n}{x}\Bigg|_{x_i}=\dv[l]{f}{x}\Bigg|_{x_i}
\end{equation}
for $N$ data points $x_i$. Thus the limited Taylor coefficient information
is buttressed by constraints asking derivatives of the 
rational approximation to match
the Taylor coefficients at multiple points $x_i$.
In practice, if we have enough data points and a high enough
order of the Taylor series, 
i.e. when $m+n+1=Ns$, \equatref{eq:mpp} yields a system
of equations that can be solved to obtain the coefficients
$a_i$ and $b_i$. When $m+n+1<Ns$, one can fit the rational
function coefficients, obtaining them through 
a maximum likelihood approach.

Unfortunately even less seems to be known about the multi-point Padé.
To ensure that rational functions constructed in this way carry useful
and robust information, we have carried out many numerical experiments.
For example in the context of the 1-$d$ Thirring model, it was shown the
multi-point Padé captures the exact chiral condensate
well~\cite{DiRenzo:2020cgp} and outperforms an ordinary Padé
for capturing the exact number density~\cite{Dimopoulos:2021vrk}.
In the context of the 2-$d$ Ising model, the multi-point Padé
was successfully employed to reproduce correct critical exponents
and Lee-Yang finite size scaling~\cite{DiRenzo:2023xeg,Singh:2023bog}.
Most relevant to the QCD phase diagram, this multi-point Padé
approach was used to find the Roberge-Weiss transition
temperature\footnote{We are currently updating our
continuum-limit estimate of $\TRW$~\cite{christian}.} \cite{Dimopoulos:2021vrk} in agreement
with earlier determinations~\cite{Bonati:2016pwz,Cuteri:2022vwk},
and $\Re\mu_B$ obtained from this approach yields
Lee-Yang finite size scaling that is consistent with
expectations~\cite{francesco}.

Since the QCD CEP is expected to be in the 3-$d$, $\Z_2$ universality class,
$\beta\delta\approx1.56$. Unfortunately the exact mapping 
from $T$ and $\mu_B$ to the Ising model is not yet known.
Sufficiently close to the CEP, a linear ansatz
\begin{equation}\begin{aligned}
t &=\alpha_t\Delta T+\beta_t \Delta\mu_B\\
h &=\alpha_h\Delta T+\beta_h \Delta\mu_B,
\end{aligned}\end{equation}
where $\Delta T\equiv T-\TCEP$ and $\Delta\mu_B\equiv\mu_B-\muCEP$,
should be reliable. This suggests~\cite{Stephanov:2006dn}
\begin{equation}\label{eq:CEPfit}
\mu_{\text{LY}}=\muCEP+ c_1\Delta T+ic_2|z_c|^{-\beta\delta}\Delta
T^{\beta\delta} + c_3\Delta T^2+\order{\Delta T^3}.
\end{equation}
Our strategy is then as follows: For each temperature ensemble, we use 
multi-point Padé approximations to find the signature of the closest
singularity of $\chi_1^B$ to the origin in the complex $\mu_B$ plane.
We fit the imaginary part of the singularity according to
\equatref{eq:CEPfit}, and since phase transitions exist at
real $\mu_B$, the point where this fit crosses the $T$-axis yields
a $\TCEP$ estimate. At the same time, we fit the real part of the
singularity to \equatref{eq:CEPfit}; plugging $\TCEP$ into this fit
yields $\muCEP$.

\section{Computational set up}

We generate $36^3 \times 6$ lattices with $N_f=2+1$ HISQ quarks~\cite{Follana:2006rc} 
using \simulat~\cite{Altenkort:2021cvg,HotQCD:2023ghu}.
We select bare parameters to maintain constant physical pion mass, i.e., $m_s/m_l=27$. To overcome the sign problem, we conduct our simulations at pure imaginary chemical potentials.
For simplicity, we set $\mu_l=\mu_s$. 
We generated a varying number of configurations (ranging from 1800 to 16000) at temperatures 166.6, 157.5, 145.0, and 136.1 MeV. Configurations are spaced by either 10 or 5 molecular dynamics time units, depending on the temperature.
The HotQCD [4,4]-Padé chemical potentials~\cite{Bollweg:2022rps} are computed
from their parameters $c_{6,2}$ and $c_{8,2}$ along with the observables 
$\chi_2^B$ and $\chi_4^B$. Error bars on these chemical potentials are
obtained by a Gaussian bootstrap of size 24000 implemented in 
the AnalysisToolbox~\cite{toolbox}, which
is also used to carry out the fits.

\begin{table}
\begin{tabularx}{\linewidth}{lRRRR}
\hline\hline
 Year & Method & $\TCEP$ [MeV] & $\muCEP$ [MeV] & $\muCEP/\TCEP$   \\
 \hline
2023 & CP+LQCD~\cite{Basar:2023nkp} & $\approx100$ & $\approx580$ & $\approx5.8$ \\ 
2022 & LQCD~\cite{Bollweg:2022rps} & &  &$\gtrsim3$\\
2021 & DSE~\cite{Gunkel:2021oya} & 117 & 600 & 5.13 \\
2021 & DSE~\cite{Gao:2020fbl} & 109 & 610 & 5.59 \\
2020 & fRG~\cite{Fu:2019hdw} & 107 & 635 & 5.54 \\
2017 & BHE~\cite{Critelli:2017oub} & 89 & 724 & 8.1 \\ 
\hline\hline
\end{tabularx}
\caption{Some results and constraints on the CEP location from
conformal Pad\'e (CP) and lattice QCD (LQCD) 
along with
Dyson-Schwinger equations (DSE),
functional renormalization group (fRG), and black hole engineering (BHE).}
\label{tab:cepStatus} 
\end{table}

\begin{figure}
\centering
\includegraphics[width=0.85\linewidth]{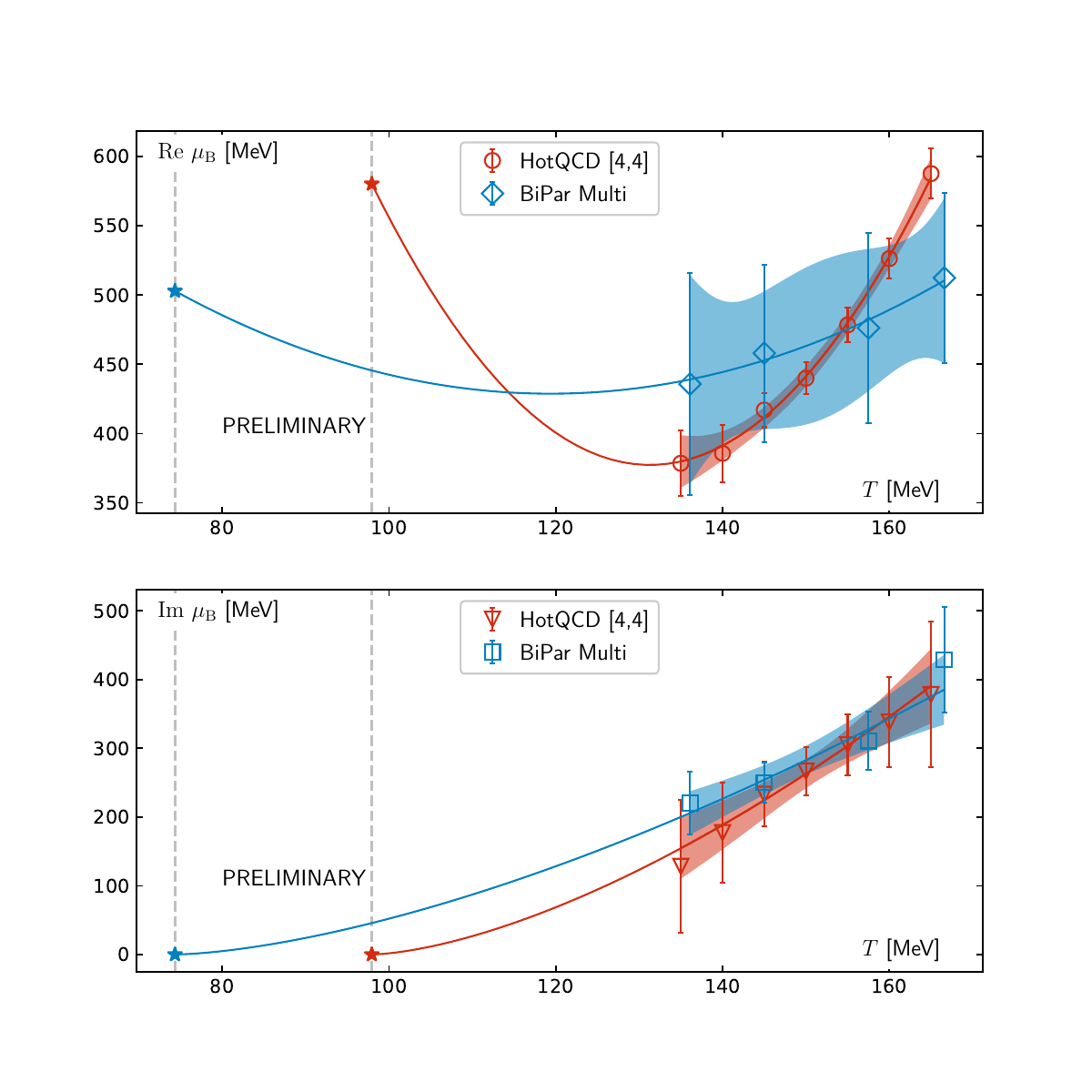}
\caption{Projections of the location of $\muCEP$ and $\TCEP$ using
the HotQCD [4,4] Padé and the present study's multi-point Padé.
Data are fit using \equatref{eq:CEPfit}.
In the lower panel, the stars indicate where $\Im\muCEP=0$,
which estimates where the critical point should be.
Following the dotted grey line upwards sets the cutoff
for the fit to $\Re\mu_B$, whose corresponding star
indicates our estimate for $\muCEP$.}
\label{fig:CEP}
\end{figure}

\section{Results}

In \figref{fig:CEP} we show the current status of our $(\TCEP,\muCEP)$ estimate.
The red data indicate singular $\mu_B$ estimated from HotQCD data that
used a [4,4] Padé on $N_\tau=8$ ensembles. Moreover the HotQCD data
used aspect ratio $N_\sigma/N_\tau=4$. Singular $\mu_B$ coming from
the multi-point Pad\'e are in blue. To one significant digit,
extrapolations indicate
approximate critical endpoints $(\TCEP,~\muCEP)$ of
(100~MeV,~600~MeV) and (70~MeV,~500~MeV)
for the HotQCD and Bielefeld-Parma data, respectively.
We currently do not have a strong reason to favor one estimate over the
other, hence we average these quantities, yielding approximately
\begin{equation}
(\TCEP, \muCEP)\approx(85~\text{MeV},550~\text{MeV})
~~~~\text{and}~~~~
\muCEP/\TCEP\approx 6.5.
\end{equation}
 
In \tabref{tab:cepStatus} we collect several other estimates and
constraints on the critical point from lattice data and effective models.
Our preliminary result is in rough agreement with these other approaches,
which consistently seem to find a temperature around 90-120~MeV
and a chemical potential around 500-700~MeV. All approaches seem to be
sensitive to some physical phenomenon there.

We stress that these results are still preliminary. 
The fit to singular $\Re\mu_B$ obtained from the multi-point Padé currently
suffers from having only 4 data points to fit 3 parameters. The fit should
hence stabilize with lower temperature data, which
is also crucial for the imaginary part, giving us more information on
the approach to the $T$-axis.
Our error analysis is still undergoing some refinement. Hence the
picture we give in \figref{fig:CEP} is subject to change\footnote{For
example, estimates and error bars reported here
are a refinement of those we reported
in Ref.~\cite{Goswami:2024jlc}.} slightly
as we converge on our final estimate.

\section{Summary and outlook}

The multi-point Pad\'e has been tested in a variety of situations,
including for the Ising model, Thirring model, and near the 
Roberge-Weiss critical point, which gives us confidence in the approach.
We observe a possible indication of the QCD CEP around $T\sim85$ MeV, 
$\mu_B\sim550$ MeV when $\mu_Q=\mu_S=0$.
We stress that, as of these proceedings, this result is preliminary.
In the short term, we are adding a lower temperature point
and refining our CEP estimate strategy. A long-term goal is an
eventual continuum-limit extrapolation. Lower temperature ensembles
are being generated to get a better handle on the CEP fit.
Besides probing the QCD CEP, we are also investigating the chiral 
transition with this approach.

\section*{Acknowledgements}

DAC was supported by the National Science Foundation under Grant PHY20-13064. CS and SS acknowledge support by the \textit{Deutsche Forschungsgemeinschaft} (DFG, German Research Foundation) under grant 315477589-TRR
211 and project number 460248186 (PUNCH4NFDI).

\bibliographystyle{JHEP}
\bibliography{bibliography}

\end{document}